\documentclass[sigconf]{acmart}

\usepackage{adjustbox}

\AtBeginDocument{%
  \providecommand\BibTeX{{%
    \normalfont B\kern-0.5em{\scshape i\kern-0.25em b}\kern-0.8em\TeX}}}

\setcopyright{rightsretained}

\copyrightyear{2024} 
\acmYear{2024}
\setcopyright{rightsretained}

\begin{document}

\title[Exploring Intent-Based User Interfaces through Abstract-to-Detailed Task Transitions]{Frontend Diffusion: Exploring Intent-Based User Interfaces through Abstract-to-Detailed Task Transitions}

\author{Qinshi Zhang}
\email{qiz065@ucsd.edu}
\affiliation{
    \institution{University of California, San Diego}
    \city{San Diego}
    \state{California}
    \country{USA}
}

\author{Latisha Besariani Hendra}
\email{lbesarian2-c@my.cityu.edu.hk}
\affiliation{
    \institution{City University of Hong Kong}
    \city{Hong Kong}
    \country{China}
}

\author{Mohan Chi}
\email{chi70@purdue.edu}
\affiliation{
    \institution{Purdue University}
    \city{West Lafayette}
    \state{Indiana}
    \country{USA}
}

\author{Zijian Ding}
\email{ding@umd.edu}
\authornote{Corresponding author}
\affiliation{
    \institution{University of Maryland, College Park}
    \city{College Park}
    \state{Maryland}
    \country{USA}
}

\renewcommand{\shortauthors}{Zhang, et al.}

\begin{abstract}
The emergence of Generative AI is catalyzing a paradigm shift in user interfaces from command-based to intent-based outcome specification. In this paper, we explore abstract-to-detailed task transitions in the context of frontend code generation as a step towards intent-based user interfaces, aiming to bridge the gap between abstract user intentions and concrete implementations. We introduce Frontend Diffusion, an end-to-end LLM-powered tool that generates high-quality websites from user sketches. The system employs a three-stage task transition process: sketching, writing, and coding. We demonstrate the potential of task transitions to reduce human intervention and communication costs in complex tasks. Our work also opens avenues for exploring similar approaches in other domains, potentially extending to more complex, interdependent tasks such as video production.

\end{abstract}



\begin{teaserfigure}
  \includegraphics[width=1\textwidth]{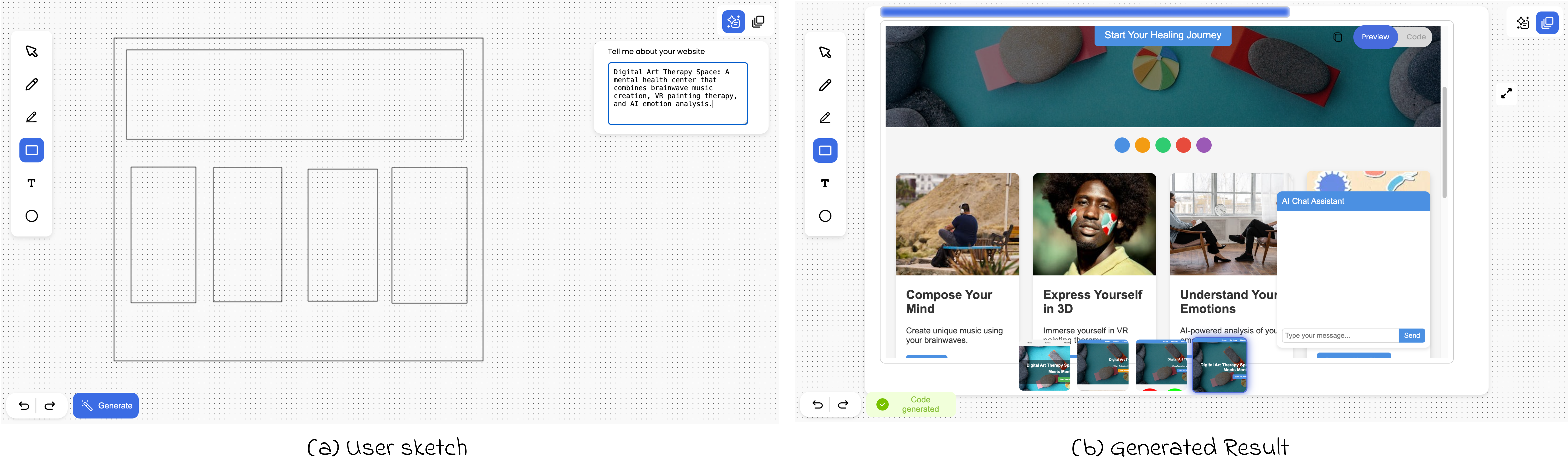}
  \caption{Frontend Diffusion is an end-to-end, LLM-powered tool for high-quality frontend code generation. Users can input a rough sketch on the canvas (left), which the system then transforms into a high-quality website page (right).}
 \Description{Frontend Diffusion.}
  \label{fig:teaser}
\end{teaserfigure}

\maketitle

\section{Introduction}

The development of Generative AI, particularly the capabilities of Large Language Models (LLMs) in interpreting and executing natural language, may be viewed as heralding the first new user interface paradigm shift in 60 years \cite{nielsen2023ai}. This shift moves from command-based interactions, typified by command line interfaces and graphical user interfaces, to intent-based outcome specification \cite{nielsen2023ai}. This emerging intent-based paradigm potentially enables users to communicate their intentions to machines without necessarily translating them into machine-comprehensible commands, whether through programming languages or graphical buttons. This shift may foster interfaces that support more abstract human expressions, especially for command-intensive tasks such as coding \cite{dingAdvancingGUIGenerative2024,ding2024towards}.

Currently, the interfaces for command-intensive tasks continue to necessitate substantial human intervention, where individuals typically specify incremental steps while AI generates corresponding code, akin to agile programming \cite{zhu-tianSketchThenGenerate2024}. However, ongoing advancements in Generative AI capabilities suggest the potential for developing a framework that may bridge the gap between intent-level expression and command-level implementation, potentially enhancing output quality while reducing the need for extensive human intervention. Previous research has demonstrated that Generative AI, such as Large Language Models (LLMs), can complete fixed-scope content curation tasks based on human intent without further intervention or intent iteration. For example, LLMs have shown promise in text summarization tasks \cite{dingHarnessingPowerLLMs2023}. However, Generative AI require greater human intervention for tasks involving increasing amounts of information \cite{dingFluidTransformersCreative2023, dingMappingDesignSpace2023}.
It motivates us to develop more effective scaffolding paradigm for Generative AI to respond to human intent and complete tasks in an agent-like manner.

\begin{figure}[h]
\centering
\includegraphics[width=1\linewidth]{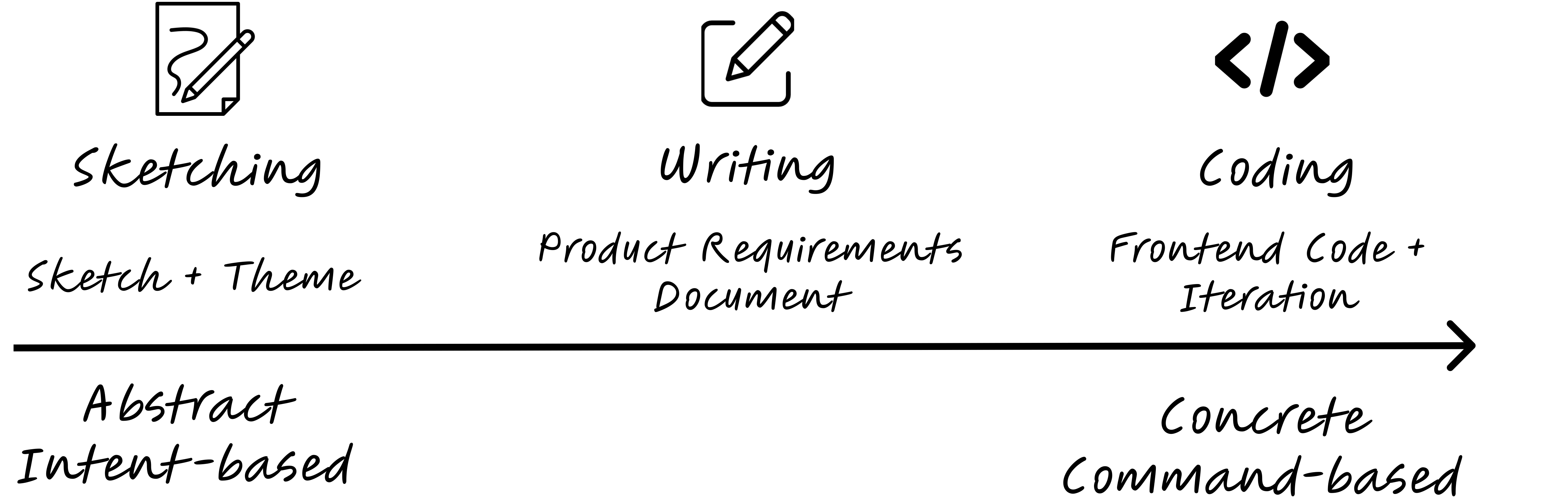}
    \caption{Task transition paradigm to bridge more abstract, intent-based task expression to more concrete, command-based implementations. The task transition process flows from sketching (sketch and theme inputs) to writing (product requirement document generation) and coding (frontend code generation and iteration).}
    \label{fig:spectrum}
    \Description{Spectrum.}
\end{figure}

Recent research has indicated the feasibility of bridging intent expression in abstract tasks to concrete implementation at a more granular level. Examples include the transition from sketching to writing \cite{chungTaleBrushVisualSketching2022} and from design to data analysis \cite{ding2024intelligent}. Building upon these findings, we propose exploring more extensive intent-to-command transitions, such as progressing from sketching to writing (planning) and ultimately to coding (see Figure \ref{fig:spectrum}). Our choice of website frontend generation as a user interface coding task \cite{wuUICoderFinetuningLarge2024} is motivated by its similarity to sketching. In both cases, the code or sketch serves as a representation of visual elements \cite{siDesign2CodeHowFar2024}.

\section{System Design}

We introduce Frontend Diffusion, an end-to-end LLM-powered high-quality frontend code generation tool, spanning from sketching canvas to website previews. As outlined in the introduction, the frontend generation task progresses through three stages: sketching, writing, and coding Our system utilizes the Claude 3.5 Sonnet language model (Sonnet)\footnote{https://www.anthropic.com/news/claude-3-5-sonnet} for all text and code generation. While Claude represents one of the most advanced language models as of July 2024, we anticipate rapid developments in Generative AI. Therefore, the task transition techniques described herein are designed to be model-agnostic, ensuring their applicability to future, more advanced Generative AI models.

\subsection{Sketching: Visual Layout Design and Theme Input}
The system's initial phase comprises a graphical user interface with two key components: a canvas panel for visual representation of the envisioned website layout, and a prompt panel for textual descriptions of the website theme. Upon completion of the user's sketch and theme input, the user can activate the code generation process via "Generate" button. The system then converts the sketch into SVG format, followed by a subsequent transformation into JPG format. This two-step conversion process was implemented based on empirical evidence from our tests, showing that language models exhibit better performance when processing images in JPG format compared to images in SVG format. 

\subsection{Writing: Product Requirements Document Generation}
This phase transforms the user's visual and textual inputs into a structured document, referred to as the Product Requirements Document (PRD), which serves as a blueprint for the website's development process. The PRD generation process leverages Sonnet. To enhance the visual appearance of the generated websites, the system integrates the Pexels API\footnote{https://www.pexels.com/api/} for image retrieval. The language model is specifically prompted to include image terms and size descriptions (e.g., [school(large)]). These descriptors are subsequently utilized to query the Pexels API, which returns relevant image URLs for incorporation into the PRD.

\subsection{Coding: Website Generation and Iterative Refinement}
The coding phase of the system consists of two primary components: (1) Initial code generation:
the system utilizes the generated PRD and the original user prompt as inputs for code generation, employing Sonnet to produce the initial website code; (2)  Iterative refinement: the system implements an iterative refinement process to automatically enhance the generated website with richer functionality and reduced flaws. This process involves analyzing the initial code to generate optimization suggestions, merging these suggestions with the original theme, and utilizing the enhanced theme along with the previously generated PRD to regenerate the code. The system executes this iterative refinement process multiple times (by default, n=4). Users can navigate between iterations by selecting preview thumbnails displayed at the interface's bottom, and can access or copy the generated code for each version.

\begin{figure*}[htbp]
    \centering
\includegraphics[width=0.81\linewidth]{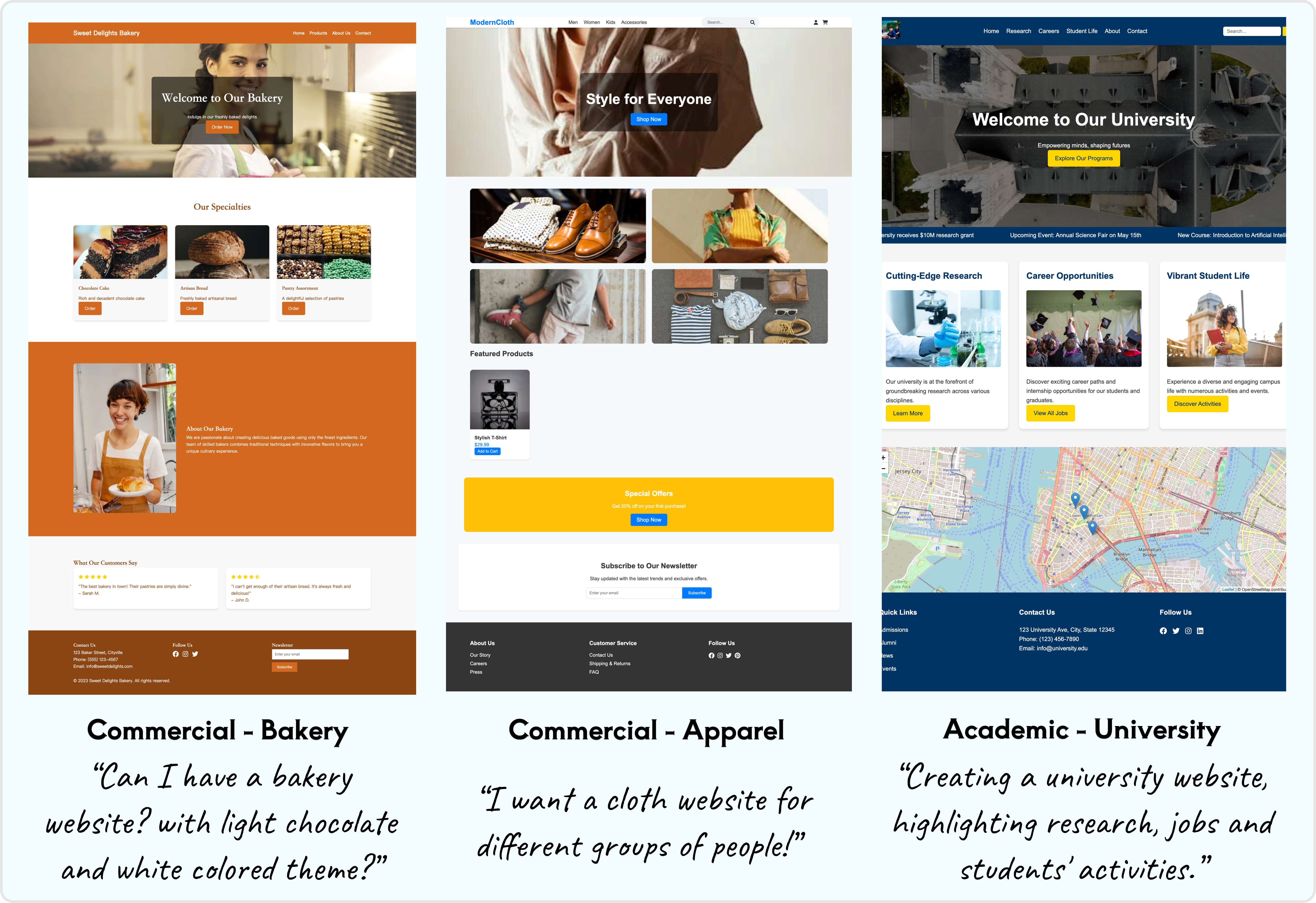}
    \caption{Screenshots of generated websites spanning commercial and academic domains. The theme prompt as user intent to generate each website is shown beneath the corresponding website.}
    \label{fig:examples}
    \Description{Examples.}
\end{figure*}

\section{Results and Discussion}

The generated websites, as illustrated in Figure \ref{fig:examples}, exhibit generally satisfactory visual appearances. These include contextually appropriate textual content, imagery, color schemes, layouts, and functionalities. Those results align with our "intent-based" objective of only requiring users to express their intent and scaffolding Generative AI to deliver the final output, potentially reducing the communication costs between users and Generative AI systems.

This task transition paradigms may motivate further exploration of intent-based interfaces, potentially extending to more complex tasks with interdependent components such as video generation. For example, we might envision an abstract-to-detailed task transition process for generating video advertisements that begins with sketches and thematic inputs, transitions to script writing, then proceeds to generate textual and visual descriptions of storyboards, followed by video generating end editing, and culminating in iterative video refinement. We aim to further investigate the potential of intent-based user interfaces in streamlining complex, interdependent workflows across various domains.

Future work could focus on studies empirically validating the effectiveness of this task transition approach in more diverse and complex task environments. Additionally, research into optimizing the task transition process and enhancing the quality of inter-task communication may yield improvements in the overall performances.

\bibliographystyle{ACM-Reference-Format}
\bibliography{sample-base}


\end{document}